\documentclass[10pt, twocolumn,conference]{IEEEtran}
\IEEEoverridecommandlockouts

\usepackage{tabto}
\usepackage{algpseudocode}
\usepackage{algorithm}
\usepackage{indent}
\usepackage{cite}
\usepackage{amsmath,amssymb,amsfonts}
\usepackage{graphicx}
\usepackage{textcomp}
\usepackage{xcolor}
\def\BibTeX{{\rm B\kern-.05em{\sc i\kern-.025em b}\kern-.08em
    T\kern-.1667em\lower.7ex\hbox{E}\kern-.125emX}}
\begin{document}

\title{Hybridizing Base-Line 2D-CNN Model with Cat Swarm Optimization for Enhanced Advanced Persistent Threat Detection
\thanks{}
}

\author{\IEEEauthorblockN{Ali M. Bakhiet}
\IEEEauthorblockA{\textit{Computer Science Department } \\
\textit{Culture and Science October 6th City}\\
Giza, Egypt
}
\and
\IEEEauthorblockN{Salah A. Aly}
\IEEEauthorblockA{\textit{Computer Science Section} \\
\textit{Faculty of Science, Fayoum University}\\
Fayoum, Egypt
}
}

\maketitle
\begin{abstract}
In the realm of cyber-security, detecting Advanced Persistent Threats (APTs) remains a formidable challenge due to their stealthy and sophisticated nature. This research paper presents an innovative approach that leverages Convolutional Neural Networks (CNNs) with a 2D baseline model, enhanced by the cutting-edge Cat Swarm Optimization (CSO) algorithm, to significantly improve APT detection accuracy. By seamlessly integrating the 2D-CNN baseline model with CSO, we unlock the potential for unprecedented accuracy and efficiency in APT detection. The results unveil an impressive accuracy score of $98.4\%$, marking  a significant enhancement in APT detection across various attack stages, illuminating a path forward in combating these relentless and sophisticated threats.
\end{abstract}


\section{Introduction}

In today's digitally connected world, the protection of sensitive information and critical infrastructure is paramount. As technology advances, so do the threats from those seeking to exploit vulnerabilities. Among these threats, Advanced Persistent Threats (APTs) pose a significant challenge in cyber security and information protection.

Machine learning is increasingly adopted by information security researchers and enterprises due to its potential for detecting assaults, particularly those that evade conventional signature-based intrusion detection systems. Detecting APTs is akin to finding a needle in a haystack; their activities closely mimic legitimate network traffic, rendering traditional detection methods inadequate. In this challenging landscape, the integration of cutting-edge technologies is imperative. Deep learning, with its prowess in processing vast volumes of data, emerges as a beacon of hope. However, harnessing the full potential of deep learning for APT detection requires more than just advanced neural networks; it demands the synergy of modern optimization algorithms.

Unlike conventional attacks that are easily detected by signature-based intrusion detection systems, APTs are characterized by their sophistication, persistence, and insidious nature. These stealthy adversaries operate with a singular objective – to infiltrate networks, remain undetected, and quietly exfiltrate sensitive data over an extended period. In doing so, they challenge the very foundations of cybersecurity and pose a substantial risk to the confidentiality, integrity, and availability of data.

This research introduces a pioneering approach that combines Convolutional Neural Networks (CNNs) with Cat Swarm Optimization (CSO)—an optimization algorithm inspired by the collective behavior of cats. This fusion aims to revolutionize APT detection by enhancing accuracy, reducing false positives, and improving the efficiency of cybersecurity operations. We utilize the "DAPT 2020" dataset, a meticulously constructed benchmark reflecting real-world APT scenarios. The cornerstone of our endeavor is the 2D-CNN baseline model, further enhanced through hybridization with CSO.

The paper is structured as follows: Section II presents related work. Section III describes the DAPT 2020 dataset. Section IV details our model and algorithm procedures. Section V discusses simulation studies for the proposed algorithms. Finally, Section VI concludes the paper.

\section{RELATED WORK}

Chu et al. \cite{Chu2006} introduced the first Cat Swarm Optimization (CSO) met-heuristic optimization technique in 2006. CSO, inspired by the collective behavior of cats, has shown remarkable capability in solving complex optimization problems and forms the backbone of our approach to enhancing APT detection.

A comprehensive review by the authors in \cite{Aram2020} evaluated CSO's developments and applications across various fields. CSO operates in two independent modes: seeking mode and tracing mode. The authors tested CSO on 23 classical benchmark functions and 10 modern benchmark functions, comparing its performance against three powerful optimization algorithms: Dragonfly Algorithm (DA), Butterfly Optimization Algorithm (BOA), and Fitness Dependent Optimizer (FDO).

Recently, the authors in~\cite{jaafer2022} created a dataset encompassing the entire life cycle of an APT attack, focusing on detection during different phases such as reconnaissance, initial compromise, lateral movement, and data exfiltration. They compared their classification model with traditional classifiers like random forest, decision tree, and K-nearest neighbor, achieving a detection accuracy of 99.89\% using only 12 out of 65 features in their dataset.

The study in~\cite{salim2023} reviewed 75 articles published between 2012 and 2022, examining various APT detection techniques, empirical experiment methods, and how APT malware was identified using these techniques.

Several other works have developed models for detecting APTs based on machine learning strategies ~\cite{Krishnapriya2022},[9]-[14], while others have focused on deep learning approaches to achieve better accuracy [15]-[17]. Additionally, research has explored hybrid methods to enhance detection performance [18]-[23], \cite{Linrui2024}, \cite{Kumar2023}.

\section{Model Datasets}

Datasets are essential for creating machine learning models that can identify sophisticated and complicated dangers like Advanced Persistent dangers (APT). APT-datasets, however, that may be utilized for modeling and detecting APT assaults are not yet available.

Several datasets have been proposed for network intrusion detection systems and cybersecurity, such as CICIDS 2018 \cite{cicids2017}, CICIDS 2017 \cite{UNB2015}, UNB 2015 \cite{ROC-Goadrich2006}, and others \cite{DAPT2020}.

Our research centers on the development of a 2D-CNN baseline model trained on the DAPT 2020 dataset. Recognizing the capability of CNNs to capture spatial features within network traffic data, we aim to optimize this baseline model. The DAPT 2020 dataset was created using network traffic collected over five days, with each day simulating three months of real-world traffic. This dataset is designed to help researchers understand anomalies, the relationships between different attack vectors, and hidden correlations that aid in early APT detection.

Generic intrusion datasets have three major limitations:
\begin{enumerate}
  \item They only include attack traffic at external endpoints, limiting their applicability for APTs, which also involve internal network attack vectors.
  \item They make it difficult to distinguish between normal and anomalous behavior, making them unrepresentative of the sophisticated nature of APT attacks.
  \item They lack the data balance characteristic of real-world scenarios, making them suitable for supervised models but inadequate for semi-supervised learning.
\end{enumerate}

The DAPT 2020 dataset addresses these issues by including assaults categorized as Advanced Persistent Threats (APTs). These assaults are difficult to differentiate from legitimate traffic flows when looking at the raw feature space and include both internal (private network) and public-to-private interface traffic. We benchmark the DAPT 2020 dataset on semi-supervised models, demonstrating that they perform poorly in detecting attack traffic at various stages of an APT due to substantial class imbalance.

\begin{figure}[htbp]
\centerline{\includegraphics[width=.5\textwidth, height=.4\textwidth]{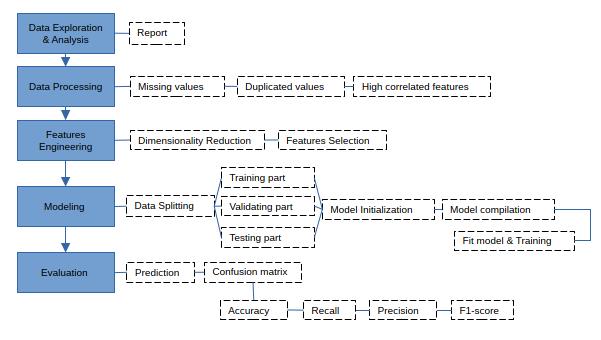}}
\caption{The proposed model framework of CSO-2D-CNN}
\label{fig}
\end{figure}

\section{MODEL AND ALGORITHM PROCEDURE}

In this section, we will demonstrate the methodology of our model.
In our methodology, we initiate by initializing N cats representing potential solutions within the optimization space. These cats undergo evaluation based on a fitness function, and the top-performing cats are stored in memory. Our model functions in two primary modes: Seeking Mode, where individual cats explore the solution space, and Tracing Mode, involving collaborative refinement of solutions among cats. To validate the efficacy of anomaly detection, we employ the CSO-2D-CNN model, as illustrated in Figure 1.

The anomaly detection approach entails training the CSO-2D-CNN model on normal network traffic data and evaluating incoming traffic packets during testing. In the training phase, the model learns baseline patterns from normal network traffic data. Subsequently, during the testing phase, incoming traffic packets traverse through the CSO-2D-CNN model, and their normalized reconstruction error is computed. If the error surpasses a predefined threshold, the packet is deemed anomalous; otherwise, it is classified as normal network traffic.


\begin{algorithm}\label{alg1}
\caption{CSO-2D-CNN Model Training and Evaluation}
\begin{algorithmic}[1]
\Require DAPT2020 dataset on semi-supervised models
\Ensure Accuracy, Loss and saved model (CSO-2D-CNN)
\State \textbf{Step 1: Model Initialization}
   \State~~-~Initialize  parameters for  2D-CNN model.
   \State~~-~Initialize  parameters for  CSO alg., including  fitness func. and  main optimiz. func.
   \State~~-~Hybridize  CSO alg. with  2D-CNN  to create  CSO-2D-CNN .
\State \textbf{Step 2: Data Preprocessing and Analysis}
   \State~~-~Perform data preprocessing  on  DAPT2020 dataset.
   \State~~-~Apply feature engineering and selection techniques tailored for  CSO-2D-CNN model.
   \State~~-~Encode  data and scale it appropriately for CSO-2D-CNN model training.
\State \textbf{Step 3: Model Training and Evaluation}
   \State~~- Train  CSO-2D-CNN model on the preproc. dataset.
   \State~~- Validate  performance of the CSO-2D-CNN model using appropriate validation techniques.
   \State~~-  Evaluate  performance of the CSO-2D-CNN model.
\State \textbf{Step 4: Model Saving}
   \State~~- Save  CSO-2D-CNN model with its trained parameters.
\State \textbf{Step 5: Documentation}
   \State~~- Document the alg. used for the CSO-2D-CNN model.
\end{algorithmic}
\end{algorithm}

\topmargin=-0.7in

\medskip

Our model development workflow encompasses several crucial steps. Initially, we perform data preprocessing tasks to prepare the dataset for model training. Next, we engineer features to capture pertinent information from the network traffic data. Following this, the dataset undergoes splitting into training and validation sets, with the CSO-2D-CNN model being trained on the training data. The trained model is then validated on the validation set to ensure generalization. Finally, we evaluate the performance of the CSO-2D-CNN model using precision scores to gauge its effectiveness in anomaly detection.

This methodology delineates our approach to harnessing Cat Swarm Optimization alongside 2D Convolutional Neural Networks for robust and efficient anomaly detection in network traffic as shown in Alg.\textbf{1}. \\

\smallskip

\textbf{Learning Rate Reduction:} Learning Rate Reduction dynamically adjusts the learning rate during training to enhance model performance. The technique operates based on the validation accuracy, where adjustments occur when improvements plateau. A patience of 2 epochs is set to prevent premature reductions, allowing the model to explore different weight configurations. Verbosity is configured at 1 to provide feedback on learning rate adjustments during training. Additionally, the learning rate is reduced by a factor of 0.5, facilitating more reliable convergence by halving the rate, enabling the optimizer to take smaller steps during gradient descent.\\

\textbf{Minimum Learning Rate:} We have specified a minimum learning rate of 0.00001. This ensures that the learning rate won't drop below this value, preventing it from becoming too small.\\

\smallskip

\textbf{Callbacks:} Callbacks are essential tools for monitoring and optimizing the training process. I've configured several callbacks to enhance the training of my baseline deep learning model:\\

\smallskip

\textbf{Reduce LR On Plateau:} This callback dynamically reduces the learning rate as needed to achieve better model convergence. It's especially useful when training reaches a plateau in terms of accuracy.\\

\smallskip

\textbf{Early Stopping:} Early stopping is set with a patience of 2 epochs. If the validation accuracy doesn't improve for two consecutive epochs, training will be halted early. This helps prevent overfitting and saves time by stopping training when further improvements are unlikely.\\

\smallskip

\textbf{Model Checkpoint:} This callback saves the model's weights to a file during training. It uses a specific naming format that includes the epoch number and validation loss. The model with the best validation accuracy will be saved as "Ann\_model\_Dense\_lab.h5," allowing you to retrieve the best-performing model after training.\\

\smallskip

\textbf{Loss Function:} sparse\_categorical\_crossentropy serves as a pivotal component in deep learning model training. You've selected "sparse\_categorical\_crossentropy" as the loss function for our baseline model, a common choice for multi-class classification tasks involving categorical (discrete) target variables. In our scenario, this loss function aptly addresses the classification of transactions into fraudulent or non-fraudulent categories (binary classification) based on our model's output. \\

\smallskip

\textbf{Optimizer:} Adam is a pivotal component in updating model parameters during training. Adam, short for Adaptive Moment Estimation, is the chosen optimizer for our model. It amalgamates the advantages of AdaGrad and RMSprop, adapting learning rates for each parameter individually. Renowned for its efficiency and rapid convergence, Adam proves advantageous in training deep neural networks, aligning with a diverse array of deep learning tasks. \\

\smallskip

\textbf{Epochs:} 5 signifies the frequency with which our model traverses the entire training dataset during training. Our configuration sets the number of epochs to 5, indicating that our model iterates over the dataset 5 times, refining its parameters to minimize the chosen loss function. The selection of epochs hinges on the problem's complexity and our model's convergence behavior. Opting for a relatively low number of epochs can be advantageous if our model converges swiftly without succumbing to overfitting.\\

\smallskip

\textbf{Batch Size:} 640 dictates the quantity of data samples processed in each forward and backward pass during a singular training iteration. With a batch size set at 640, our model processes 640 data samples before parameter updates. The chosen batch size significantly influences training speed, memory demands, and the quality of model updates. Opting for a batch size of 640 strikes a balance between computational efficiency and model convergence, representing a common choice in the field.

These configurations synergistically enhance the efficiency and efficacy of model training. Learning rate reduction facilitates convergence to an optimal solution, while callbacks mitigate overfitting and facilitate the preservation of the best-performing model for future utilization. Employing these strategies is paramount for attaining superior outcomes in machine learning endeavors such as fraud detection.

\topmargin=-0.7in

\section{RESULTS OF THE APT ATTACKS DATASET }
In this section, we will demonstrate the results and performance of our model.
We utilize the Base-Line model, a 2D-CNN model, on the numeric dataset, and optimize it with the Cat Swarm Optimization (CSO) Algorithm, resulting in the proposed CSO-2D-CNN model. Following data processing and splitting, pertinent details about the dataset are as follows: it comprises 75 features and is divided into 32 parts. These parts include a training set with a shape of (55262, 75) accounting for $70\%$ of the dataset, a validation set with a shape of (6141, 75) comprising $10\%$, and a testing set with a shape of (15351, 75) encompassing $20\%$.

For evaluating the CSO-2D-CNN model's performance, we employ four key evaluation metrics: accuracy, precision, recall, and F1 score. These metrics are derived by assessing the number of true positives (TP), false positives (FP), true negatives (TN), and false negatives (FN). Specifically, Accuracy represents the likelihood of correct predictions made by the CSO-2D-CNN model across all sample data, providing an overarching measure of model prediction accuracy. The  Accuracy measurement can be evaluated as:
\begin{equation}\label{}
  Accuracy=\frac{TP +TN}{TP +FP +TN +FN}
\end{equation}

Precision Recall (PR) represents the proportion of true attacks among all the samples detected as APT attacks, and is computed by \# (Anomalous Traffic) /\#(Dataset), where \#(•) denotes the cardinality, see \cite{DAPT2020}. The  Precision measurement can be evaluated as:

\begin{equation}\label{}
Precision=\frac{TP}{TP +FP }
\end{equation}

Recall represents the proportion of detected attacks among all the true attacks.
The  Recall measurement can be evaluated as:
\begin{equation}\label{}
Recall=\frac{TP}{TP+FN}
\end{equation}
The F1-score is calculated using Precision and Recall and represents their harmonic mean. It can assess the overall performance of the model.
\begin{equation}\label{}
F1-score=\frac{2*Precision *Recall }{Precision +Recall}
\end{equation}

The  Sensitivity measurement can be evaluated as:
\begin{equation}\label{}
Sensitivity=\frac{TP}{TP +FN}
\end{equation}

The  Specificity measurement can be evaluated as:
\begin{equation}\label{}
Specificity=\frac{TN }{TN+FP}
\end{equation}

The formula for Cohen’s kappa is calculated as:

\begin{equation}\label{}
Kappa=\frac{p_o - p_e }{1-p_e}
\end{equation}
where $p_o$ is the Relative observed agreement among raters, $p_e$ is the Hypothetical probability of chance agreement.

\begin{table}[h!]\label{table:1}
\caption{SOME INFORMATION ABOUT THE BASE-LINE MODEL}
\centering
\begin{tabular}{|c|c|c|}
  \hline \hline
  Layer (type)	& Output Shape &	Param  \\ \hline
InputLayer	& [(None, 75, 1, 1)] &	0 \\
Conv2D	& (None, 70, 1, 64)	& 448 \\
Batch Normalization&	(None, 70, 1, 64)&	256 \\
MaxPooling2D	 &(None, 35, 1, 64)	&0 \\
Conv2D	&(None, 33, 1, 64)&	12352 \\
BatchNormalization	&(None, 33, 1, 64)&	256 \\
MaxPooling2D&	(None, 17, 1, 64)	&       0 \\
Conv2D	&(None, 15, 1, 64)	& 12352 \\
BatchNormalization&	(None, 15, 1, 64)&	256 \\
MaxPooling2D&	(None, 8, 1, 64)	&0 \\
Flatten	& (None, 512)	&0 \\
Dense	& (None, 64)&	32832 \\
Dense &	(None, 32)	&2080 \\
Dense	&(None, 5)&	165 \\
  \hline
  \multicolumn{3}{|l|}{Total params: 60997 (238.27 KB)}  \\
\multicolumn{3}{|l|}{Trainable params: 60613 (236.77 KB) }\\
\multicolumn{3}{|l|}{Non-trainable params: 384 (1.50 KB)} \\
\hline
\end{tabular}
\end{table}

\begin{table} [h!]
  \centering
  \caption{PERFORMANCE REPORT }\label{}
  \begin{tabular}{|c|c|}
    \hline
    \hline
    Measurement & Value\\
    \hline
Training accuracy	& 98.8\%    \\
Validating accuracy	 & 98.2\%  \\
Testing accuracy	&  98.4\% \\
Precision Score & 	98.4\% \\
Recall Score & 	983\% \\
F1 Score	& 98.4\% \\
Sensitivity	&  99.9\% \\
Specificity & 	99.8\% \\
PPV	 &  98.9\% \\
NPV	&  99.9\% \\
Kappa Score & 	97.4\% \\
    \hline
  \end{tabular}
\end{table}

\topmargin=-0.7in

\begin{table} [h!]
  \centering
  \caption{THE CLASSIFICATION REPORT}\label{}
  \begin{tabular}{|c|c|c|c|c|}
    \hline \hline
&Precision	&Recall &	F1-score&	Support \\
\hline
Benign	&0.99	&0.98 &	0.99&	8716 \\
Data	&1.00&	1.00	&1.00	&2060 \\
Establish&	0.99&	0.98&	0.98	&1725 \\
Lateral&	0.84&	 0.94	& 0.89 &	490 \\
Reconn	&0.98	& 0.98&	0.98	&2360\\
        \hline
  \end{tabular}
\end{table}

    \begin{table} [h!]
  \centering
  \caption{ACCURACY REPORT}\label{}
  \begin{tabular}{|c|c|c|c|c|}
    \hline \hline
    accuracy	&	& &	0.98&	15351  \\
macro avg&	0.96&	0.98	&0.97&	15351  \\
weighted avg	&0.98&	0.98	&0.98	&15351  \\
    \hline
  \end{tabular}
\end{table}

The computation of the model begins with the construction of a convolutional neural network (CNN) architecture, specifically a CSO-2D-CNN model, utilizing layers including Conv2D, BatchNormalization, MaxPooling2D, Flatten, and Dense. The model's hyperparameters, such as the learning rate, batch size, and number of epochs, are tuned using the Cat Swarm Optimization (CSO) algorithm, which optimizes the model's performance through iterative exploration and exploitation steps. During optimization, the algorithm dynamically adjusts parameters to minimize the loss function, enhancing accuracy and convergence.

The CSO-2D-CNN model undergoes training on the dataset, with evaluation metrics including accuracy and loss being monitored using callbacks such as learning rate reduction and model checkpointing. Through multiple iterations, the CSO refines the model's parameters, with each iteration evaluating the fitness of the model against the validation dataset.
\smallskip

Upon completion, the best hyperparameters and fitness values are determined, showcasing the optimized CSO-2D-CNN model's effectiveness in achieving high accuracy (98.4\%) and low loss (0.048) metrics. The entire process, including model construction, hyperparameter optimization, training, and evaluation, culminates in a computational time of approximately 53 minutes.
\smallskip

After training the proposed CSO-2D-CNN model, we achieved a best fitness of (0.9835, 0.0484), signifying a remarkable accuracy of 98.35\% and a minimal loss value of 0.0484. Subsequently, the performance curves of the model post-optimization are presented below:

\begin{figure}[htbp]
\centerline{\includegraphics[width=.5\textwidth, height=.4\textwidth]{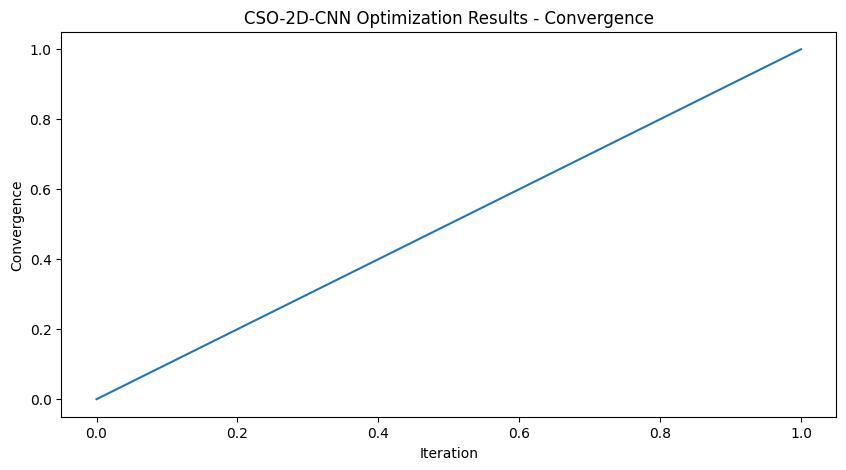}}
\caption{Convergence - Optimization Results Curves depicting the detection of attacks across various stages of an APT using the proposed CSO-2D-CNN model trained on the DAPT2020 dataset.}
\label{fig}
\end{figure}

\smallskip

\begin{figure}[ht]
\centerline{\includegraphics[width=.5\textwidth, height=.4\textwidth]{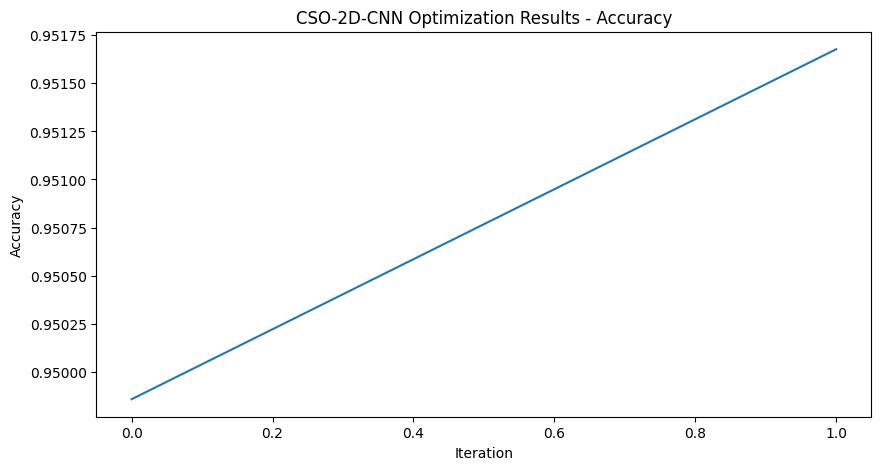}}
\caption{Accuracy - Optimization Results of CSO-2D-CNN model. Additionally, the model returns the value of the best fitness.}
\label{fig}
\end{figure}

\smallskip

\begin{figure}[htbp]
\centerline{\includegraphics[width=.5\textwidth, height=.4\textwidth]{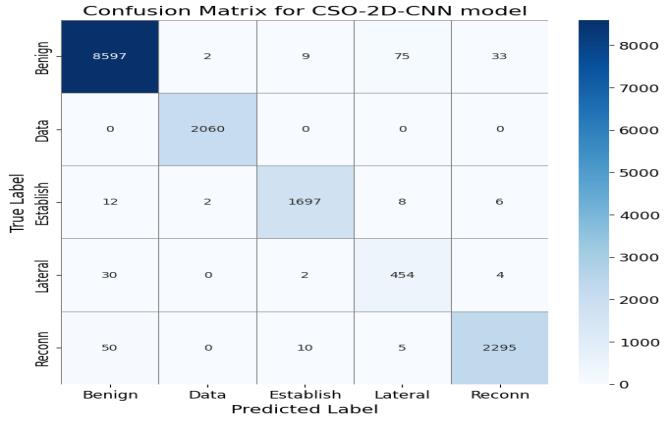}}
\caption{Confusion Matrix for CSO-2D-CNN Model}
\label{fig4}
\end{figure}


\smallskip

The confusion matrix in Figure~\ref{fig4} provides a detailed performance evaluation of the CSO-2D-CNN model. It shows  the model correctly identified 8,609 benign instances, 2,060 data attacks, 1,664 establishment attacks, 464 lateral movements, and 2,300 reconnaissance attacks. This visualization highlights the model's high accuracy in identifying benign traffic and various attack types, with minimal misclassifications. It also helps to identify specific areas for improvement, such as reducing the false positives in benign and recon categories, to further enhance the model's overall performance in APT detection.

\begin{figure}[htbp]
\centerline{\includegraphics[width=.5\textwidth, height=.4\textwidth]{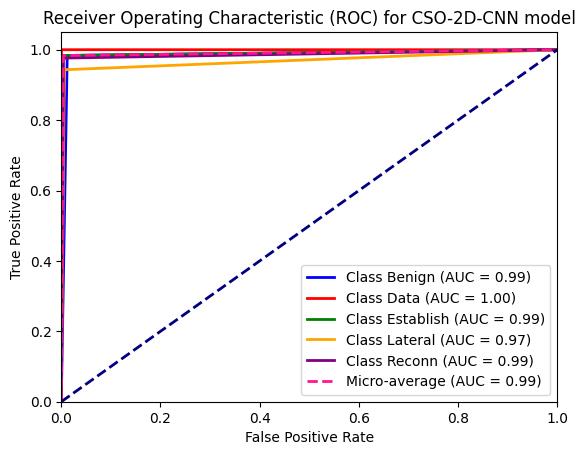}}
\caption{Receiver Operating Characteristic (ROC) curve for CSO-2D-CNN model}
\label{fig5}
\end{figure}
\smallskip

Figure~\ref{fig5} illustrates the Receiver Operating Characteristic (ROC) curve for the CSO-2D-CNN model, depicting the trade-off between true positive rate and false positive rate at various threshold settings. The ROC curve is a fundamental tool for evaluating the diagnostic ability of the model, and a higher area under the curve (AUC) indicates better performance. The ROC analysis confirms the model's strong predictive capabilities, validating the effectiveness of the hybrid CSO-2D-CNN approach in anomaly detection tasks.

\smallskip

\section{Conclusion}
In conclusion, this paper presents a pioneering framework demonstrating the symbiotic relationship between modern optimization algorithms and deep learning models for APT detection. The integration of BaseLine 2D-CNN models with Cat Swarm Optimization yields promising results, with our optimized CSO-2D-CNN model achieving an impressive accuracy of $98.35\%$ and a minimal loss value of $0.0484$ in APT detection. This underscores the significance of innovative approaches in the dynamic realm of threat detection and cybersecurity. We will also clarify the capabilities of the model before and after the improvement, elucidating the enhancements introduced by the optimization process. Such elucidation will ascertain the worthiness of the proposed model's improvement, providing insights into its efficacy in advancing APT detection capabilities.

\topmargin=-0.7in


\end{document}